\documentclass[prd,
,superscriptaddress,
nofootinbib,%
tightenlines ]{revtex4}
\usepackage{epsfig}
\usepackage[colorlinks=true,linkcolor=blue,urlcolor=blue,citecolor=blue]{hyperref}
\usepackage{amsfonts}
\usepackage{amssymb}

\newcommand{\kslash}{\not\hspace{-0.7mm}k}
\newcommand{\ben}{\begin{displaymath}}
\newcommand{\een}{\end{displaymath}}
\newcommand{\be}{\begin{equation}}
\newcommand{\ee}{\end{equation}}
\newcommand{\bea}{\begin{eqnarray}}
\newcommand{\eea}{\end{eqnarray}}

\begin{document}
\title{{ Chiral theory of} nucleons and pions in
the presence of an external gravitational field}
\author{H.~Alharazin}
\affiliation{Ruhr University Bochum, Faculty of Physics and Astronomy,
Institute for Theoretical Physics II, D-44870 Bochum, Germany}
\author{D.~Djukanovic}
 \affiliation{Helmholtz Institute Mainz, University of Mainz, D-55099 Mainz, Germany}
  \affiliation{GSI Helmholtzzentrum f\"ur Schwerionenforschung, Darmstadt, { Germany}}
\author{J.~Gegelia}
\affiliation{Ruhr University Bochum, Faculty of Physics and Astronomy,
Institute for Theoretical Physics II, D-44870 Bochum, Germany}
\affiliation{Tbilisi State  University,  0186 Tbilisi,
 Georgia}
\author{M. V.~Polyakov}
\affiliation{Ruhr University Bochum, Faculty of Physics and Astronomy,
Institute for Theoretical Physics II, D-44870 Bochum, Germany}
\affiliation{Petersburg Nuclear Physics Institute, 
		Gatchina, 188300, St.~Petersburg, Russia}

\begin{abstract}

We extend the standard second order effective chiral Lagrangian of pions and
nucleons by considering the coupling to an external gravitational field.  As an
application we calculate one-loop corrections to the one-nucleon matrix element
of the energy-momentum tensor to fourth order in chiral counting, and
{next-to-leading order tree-level} amplitude of the pion-production in an
external gravitational field.  We discuss the relation of the obtained results
to experimentally measurable observables.  Our expressions for the chiral
corrections to the nucleon gravitational form factors differ from those in the
literature. That might require to revisit the chiral
extrapolation of the lattice data on the nucleon gravitational form factors
obtained in the past.

\end{abstract}

\maketitle

\section{Introduction}	
Three basic global mechanical properties of the nucleon (mass, spin and the $D$-term\footnote{The name ``$D$-term" is rather technical, it can be traced back to more or less accidental
 notations chosen in Ref.~\cite{Polyakov:1999gs}. Nowadays, given more clear {physical} meaning of this quantity, we might call this term as ``{\it Druck-term}" derived from 
german word for pressure. })  can be obtained as a linear response of the effective action to the change of the
space-time metric\footnote{Just recall that  in classical physics our intuitive perception of the mass is related to the gravity (weighing experiment), also recall
the classical experiment  with  the Foucault pendulum to measure the { Earth's}  rotation. }.  The mass, spin and $D$-term correspond to the hadron gravitational form factors (GFFs) at zero momentum transfer \cite{Kobzarev:1962wt,Pagels:1966zza}.
While the  mass and spin of the nucleon are well-studied and well-measured quantities, the third mechanical characteristics (the $D$-term)
is more subtle, as it is related to the distribution of the internal  forces inside the nucleon \cite{Polyakov:2002yz} (for a  review see Ref.~\cite{Polyakov:2018zvc}).  The nucleon gravitational form factors are measurable experimentally 
in exclusive processes like deeply virtual Compton scattering  (DVCS) \cite{Ji:1996ek,Radyushkin:1997ki} and hard exclusive meson production \cite{Collins:1996fb}.

The first  results of  measurements
 of the $ D$-term
in hard QCD processes became available  in Refs.~\cite{Kumericki:2015lhb,Nature} for the nucleon, and in Ref.~\cite{Kumano:2017lhr} for the pion. Profound studies of all
subtleties in the extraction of the D-term  from hard exclusive processes can be found in Ref.~\cite{Kumericki:2019ddg}.
The GFFs  have been also studied in lattice QCD, see  Refs.~\cite{Shanahan:2018nnv,Shanahan:2018pib,Alexandrou:2013joa,Bratt:2010jn,Hagler:2007xi} and references therein.

The hard exclusive process can be used not only to access the GFFs of the nucleon, but also one can study other hadronic processes induced by the gravitational interaction.
For example, the pion graviproduction off  the nucleon \cite{Polyakov:1998sz,Guichon:2003ah,Polyakov:2006dd,Kivel:2004bb}. 

For systematic studies of  hadronic processes
induced by gravity in the low-energy domain one needs to derive the Effective Chiral Lagrangian (EChL) for nucleons and pions in  curved space-time.
The corresponding EChL for pions has been derived in Ref.~\cite{Donoghue:1991qv}, 
and the GFFs of the pion obtained using EChL can be found in Ref.~\cite{Kubis:1999db}. In the present work we  write down the full EChL of pions and nucleons in curved space-time up to second order. For that  we
couple the standard Lagrangian of chiral EFT up to order two \cite{Gasser:1984yg,Fettes:2000gb}  to the gravitational field and introduce two additional terms 
which depend explicitly on the curvature characteristics of the space-time.  Although these additional terms are zero in flat space-time, they contribute to the energy-momentum tensor (EMT) of pions and nucleons in Minkowski space-time and hence 
to GFFs of the nucleon as well as to hadronic processes induced by gravitational interaction.  

\medskip

In this work we apply the derived EChL to:

\begin{itemize}
\item
 manifestly Lorentz-invariant calculations of  one-loop contributions to the nucleon gravitational form factors up to fourth order according to  standard power counting rules. To remove the divergences and the power counting violating 
contributions from one-loop diagrams we apply the EOMS renormalization scheme 
of Refs.~\cite{Gegelia:1999gf,Fuchs:2003qc}. We obtain the result which is at variance  with the calculations of Ref.~\cite{Diehl:2006ya} done to the same chiral order 
 using the heavy baryon formalism \cite{Jenkins:1990jv,Bernard:1992qa}. 
The origin of this difference was clarified with the authors of Ref.~\cite{Diehl:2006ya} - they agreed with our results.
Our new expressions for the chiral corrections to nucleon GFFs 
 might require revisiting the chiral extrapolation of the lattice data on these quantities obtained in the past.
 \item
 derivation of the large distance asymptotic of the energy, spin, pressure and shear force distributions in the parametrically wide region of distance $1/\Lambda_{\rm strong}\ll r \ll 1/M_\pi$.
\item
calculation of the amplitude of the pion graviproduction to next-to-leading order.  The pion graviproduction can be accessed in hard exclusive processes \cite{Polyakov:1998sz,Guichon:2003ah,Polyakov:2006dd,Kivel:2004bb} and can be used
to get additional information about the new low-energy constants of the EChL. 

\end{itemize}
Surely, applications of the EChL derived here are not limited to the above physics problems. It can be used for a wide spectrum of applications,
ranging from the 
physics of  hadronic reactions in recently observed violent events, such as the neutron stars mergers \cite{TheLIGOScientific:2017qsa}, to the
fundamental questions of General Relativity (see, e.g., Ref.~\cite{Avelino:2019esh}), and 
to the theory of hard exclusive processes
and physics of exotic hadro-charmonia  \cite{Dubynskiy:2008mq,Eides:2015dtr,Perevalova:2016dln}.

Our paper is organized as follows: In section~\ref{effective_Lagrangian} we
 obtain the full second order EChL for pions and nucleons in curved space-time  and the corresponding expression for EMT.
Next we calculate the nucleon matrix element of the EMT in section~\ref{NFFs}.  
The large distance asymptotic of the energy, spin, pressure and shear force distributions is studied in subsection~\ref{sec:largeR}.
In section~\ref{ppr}  we discuss the  one-pion graviproduction tree-level amplitude at next-to-leading order. The results of our work are
summarized in  section~\ref{conclusions}.  The appendices contain definitions of loop integrals,  explicit expressions of GFFs in the chiral limit,   and the pion graviproduction amplitude.

\section{Effective action in curved space time and the energy-momentum tensor}
\label{effective_Lagrangian}

In this section we obtain the full second order EChL for the pions and nucleons in curved space-time  and  derive the corresponding 
expression for the EMT. The EChL for pions and nucleons without including the coupling to gravitational fields can be found in Refs.~\cite{Gasser:1984yg,Fettes:2000gb}. 
For the purpose of obtaining the EMT corresponding to these effective Lagrangians, analogously to Ref.~\cite{Donoghue:1991qv},  we consider their coupling to gravitational fields.   
The action corresponding to the leading order effective Lagrangian of pseudoscalar mesons interacting with the gravitational field is given by  \cite{Donoghue:1991qv} 
  \begin{equation}
S_{\rm \pi} = \int d^4x \sqrt{-g}\, \left\{ \frac {F^2}{4}\, g^{\mu\nu}\, {\rm Tr} ( D_\mu U  (D_\nu U)^\dagger ) + \frac{F^2}{4}\,{\rm Tr}(\chi U^\dagger +U \chi^\dagger) \right\},
\label{PionAction}
\end{equation}
where $\chi= 2 B_0(s+i p)$, $D_\mu U=\partial_\mu U -i r_\mu U +i U l_\mu $ and the $2\times 2$ unitary matrix $U$ represents the pion field. The parameter $B_0$ is related to the 
vacuum condensate and $s$, $p$, $l_\mu =v_\mu-a_\mu $ and $r_\mu =v_\mu + a_\mu $ 
are external sources.

 For the action corresponding to the leading- and next-to-leading order effective Lagrangians of nucleons interacting with pions and the gravitational field we obtain:
\begin{eqnarray}
S_{\rm \pi N} & = & \int d^4x \sqrt{-g}\, \biggl\{
\frac{1}{2} \, \bar\Psi \, i e^\mu_a\gamma^a \nabla_\mu \Psi -\frac{1}{2} \, \nabla_\mu \bar\Psi
\, i e^\mu_a\gamma^a\Psi  -m \bar\Psi\Psi +\frac{g_A}{2}\, \bar\Psi e^\mu_a\gamma^a \gamma_5 u_\mu \Psi  \nonumber\\
&+&  c_1 \langle \chi_+\rangle  \bar\Psi  \Psi  - \frac{c_2}{8 m^2} g^{\mu\alpha} g^{\nu\beta} \langle u_\mu u_\nu\rangle  \left( \bar\Psi \left\{ \nabla_\alpha, \nabla_\beta\right\}  \Psi+
 \left\{ \nabla_\alpha, \nabla_\beta\right\}  \bar\Psi \Psi \right) + \frac{c_3}{2} \, g^{\mu\nu} \langle u_\mu u_\nu\rangle  \bar\Psi  \Psi  \nonumber\\
&+&  \frac{i c_4}{4} \, \bar\Psi \,e^\mu_a e^\nu_b \sigma^{ab}\,\left[ u_\mu ,u_\nu\right] \Psi 
+ c_5 \bar\Psi\hat \chi_+ \Psi + \frac{c_6}{8m} \bar\Psi \,e^\mu_a e^\nu_b \sigma^{ab} F^+_{\mu\nu} \Psi + \frac{c_7}{8m} \bar \Psi \,e^\mu_a e^\nu_b \sigma^{ab} 
\langle F^+_{\mu\nu} \rangle \Psi  \nonumber\\
&+&  \frac{c_8}{8}\, R \bar\Psi \Psi  + \frac{ i c_9}{m} \, R^{\mu\nu} \left( \bar\Psi e_\mu^a \gamma_a \nabla_\nu  \Psi 
- \nabla_\nu  \bar\Psi e_\mu^a \gamma_a  \Psi \right)
\biggr\} ,
\label{MAction}
\end{eqnarray}
 As usual, the action at this chiral order has been reduced to the above (minimal) form by using field redefinitions.
In Eq.~(\ref{MAction}) $g^{\mu\nu}$  and $e^\mu_a$ are the metric (we use  the signature $(+,-,-,-)$) and vielbein gravitational fields, respectively,
\begin{eqnarray}
u_\mu & = & i \left[ u^\dagger \partial_\mu u  - u \partial_\mu u^\dagger -i (u^\dagger v_\mu u - u v_\mu u^\dagger )\right]\,,\nonumber\\
F_{\mu\nu}^+ &=& u^\dagger F_{R\mu\nu} u+u F_{L\mu\nu} u^\dagger \,,\nonumber\\
F_{R\mu\nu} &=& \partial_{\mu} r_\nu - \partial_\nu r_\mu-i[r_\mu,r_\nu] \,, \nonumber\\
F_{L\mu\nu} &=& \partial_{\mu} l_\nu - \partial_\nu l_\mu-i[l_\mu,l_\nu] \,,  \nonumber\\
\chi_+ & = & u^\dagger \chi u^\dagger+u \chi^\dagger u\,,\nonumber\\
\hat \chi_+ & = & \chi_+ -\frac{1}{2} \langle \chi_+\rangle\,,
\label{adddefs}
\end{eqnarray}
and the covariant derivative acting on the nucleon field has the form 
\begin{eqnarray}
\nabla_\mu \Psi &=& \partial_\mu\Psi +\frac{i}{2} \, \omega^{ab}_\mu \sigma_{ab} \Psi + \left( \Gamma_\mu  -i v_\mu^{(s)}\right)\Psi, \nonumber\\
\nabla_\mu \bar\Psi &=& \partial_\mu\bar\Psi -\frac{i}{2} \, \bar\Psi \, \sigma_{ab} \, \omega^{ab}_\mu - \bar\Psi \left( \Gamma_\mu  -i v_\mu^{(s)}\right) ,
\label{CovD}
\end{eqnarray}
where $v_\mu^{(s)}$ is an iso-scalar external vector source, $\sigma_{ab}=\frac{i}{2}[\gamma_a, \gamma_b]$ and 
\begin{eqnarray}
\Gamma_\mu & = & \frac{1}{2} \left[ u^\dagger \partial_\mu u  +u \partial_\mu u^\dagger -i (u^\dagger v_\mu u+u v_\mu u^\dagger )\right]~,\nonumber\\
\omega_\mu^{ab} &=& -g^{\nu\lambda} e^a_\lambda \left( \partial_\mu e_\nu^b
- e^b_\sigma \Gamma^\sigma_{\mu \nu} \right),\nonumber\\
\Gamma^\lambda_{\alpha \beta} &=& \frac{1}{2}\,g^{\lambda\sigma} \left( \partial_\alpha g_{\beta\sigma}
+ \partial_\beta g_{\alpha\sigma} -  \partial_\sigma g_{\alpha\beta} \right)~, \nonumber\\
R^\rho_{~\sigma\mu\nu} &=& \partial_\mu \Gamma^\rho_{\nu \sigma} -\partial_\nu \Gamma^\rho_{\mu \sigma} + \Gamma^\rho_{\mu \lambda}  \Gamma^\lambda_{\nu \sigma} - \Gamma^\rho_{\nu \lambda}  \Gamma^\lambda_{\mu \sigma}   \,,\nonumber\\ 
R &=& g^{\mu\nu} R^\lambda_{~\mu\lambda\nu}\,.
\label{omega}
\end{eqnarray}
The vielbein fields satisfy the following relations:
\begin{eqnarray}
&& e^a_\mu e^b_\nu \eta_{ab}=g_{\mu\nu}, \ \ \  e_a^\mu e_b^\nu \eta^{ab}=g^{\mu\nu}, \nonumber\\
&& e^a_\mu e^b_\nu g^{\mu\nu}=\eta^{ab}, \ \ \  e_a^\mu e_b^\nu g_{\mu\nu}=\eta_{ab}.
\label{VRels}
\end{eqnarray}
When calculating $\left\{ \nabla_\alpha, \nabla_\beta\right\}  \Psi$ and $\left\{ \nabla_\alpha, \nabla_\beta\right\}  \bar \Psi$ in Eq.~(\ref{MAction}) we need to take into account that the purely gravitational covariant derivative $\nabla^{\rm gr}_\mu$ acting on a tensor has the form:
\begin{equation}
\nabla^{\rm gr}_\mu T^{\alpha\beta\ldots}_{\rho\sigma\ldots} = \partial_\mu T^{\alpha\beta\ldots}_{\rho\sigma\ldots}  +\Gamma^\alpha_{\mu\nu} T^{\nu\beta\ldots}_{\rho\sigma\ldots} + \ldots  - \Gamma^\nu_{\mu\rho}  T^{\alpha\beta\ldots}_{\nu\sigma\ldots} -\ldots  \,.
\label{CovDT}
\end{equation}

\medskip
The effective action  of Eq.~(\ref{MAction})  contains low-energy constants (LECs),  $c_{1- 7}$ corresponding to the constants  of the second order $\pi N$
effective action introduced in Ref.~\cite{Fettes:2000gb},  the values of which are constrained by data on low-energy  physics of pions and nucleons.  
It also  contains two new LECs,  $c_{8}$ and $c_{9}$.\footnote{{ Notice that these couplings are not the same as introduced in the effective Lagrangian with external tensor sources in Ref.~\cite{Dorati:2007bk}}.} 
We shall see below that they can be constrained by GFFs of the nucleon and/or by the pion graviproduction.
It is important that these new LECs  (like all others) are universal -- the same constants enter  various hadronic processes induced by gravity. 

\medskip

 Using the definition of the EMT for matter fields interacting with the gravitational metric fields,
\begin{eqnarray}
T_{\mu\nu} (g,\psi) & = & \frac{2}{\sqrt{-g}}\frac{\delta S_{\rm m} }{\delta g^{\mu\nu}}\,,
\label{EMTMatter}
\end{eqnarray}
from the action of Eq.~(\ref{PionAction}) we obtain  { in flat spacetime}
\begin{eqnarray}
T^{{  (\pi)}}_{\mu\nu} & = &  \frac {F^2}{4}\, {\rm Tr} ( D_\mu U  (D_\nu U)^\dagger +D_\nu U  (D_\mu U)^\dagger ) 
- \eta_{\mu\nu} \left\{ \frac {F^2}{4}\, {\rm Tr} ( D^\alpha U  (D_\alpha U)^\dagger ) +  \frac{F^2}{4}\,{\rm Tr}(\chi U^\dagger +U \chi^\dagger) 
\right\} ,
\label{PionEMT}
\end{eqnarray}
where $\eta_{\mu\nu}$ is the Minkowski metric tensor. 
{ For the fermion fields interacting with { gravitational} vielbein fields we use { the definition \cite{Birrell:1982ix} }
\begin{eqnarray}
T_{\mu\nu}  (g,\psi) & = & \frac{1}{2 e} \left[ \frac{\delta S }{\delta e^{a \mu}} \,e^{a}_\nu + \frac{\delta S }{\delta e^{a \nu}} \,e^{a}_\mu  \right] ,
\label{EMTfermion}
\end{eqnarray}
where $e$ is the determinant of $e^a_\mu$. 
The action of Eq.~(\ref{MAction}) leads to the following expression for the EMT  in flat spacetime:
\begin{eqnarray}
T^{{  (\pi N)}}_{\mu\nu} & = &  \frac{i}{4} \, \left( \bar\Psi \,  \gamma_\mu D_\nu \Psi + \bar\Psi \,
\gamma_\nu D_\mu \Psi  - D_\mu \bar\Psi \, \gamma_\nu\Psi
- D_\nu \bar\Psi \, \gamma_\mu\Psi   \right)  
+ \frac{g_A}{4} \, \left( \bar\Psi \,  \gamma_\mu \gamma_5 u_\nu \Psi + \bar\Psi \,
\gamma_\nu\gamma_5 u_\mu \Psi  \right)
\nonumber\\
&-&  \frac{c_2}{8 m^2} \left[  \langle u_\mu u^\beta\rangle  \left( \bar\Psi \left\{ D_\nu, D_\beta\right\}  \Psi+
 \left\{ D_\nu, D_\beta\right\}  \bar\Psi \Psi \right) +   
 \langle u^\alpha u_\mu\rangle  \left( \bar\Psi \left\{ D_\alpha, D_\nu\right\}  \Psi+
 \left\{ D_\alpha, D_\nu\right\}  \bar\Psi \Psi \right) \right. \nonumber\\
& + & \left.  \langle u_\nu u^\beta\rangle  \left( \bar\Psi \left\{ D_\mu, D_\beta\right\}  \Psi+
 \left\{ D_\mu, D_\beta\right\}  \bar\Psi \Psi \right) +   
 \langle u^\alpha u_\nu\rangle  \left( \bar\Psi \left\{ D_\alpha, D_\mu\right\}  \Psi+
 \left\{ D_\alpha, D_\mu\right\}  \bar\Psi \Psi \right)
  \right] \nonumber\\
  & +&  \frac{i c_2}{16 m^2} \, \partial^{{ \rho}} \left\{  \langle u^\alpha u^\beta\rangle  
   \left[
 D_\alpha  \bar\Psi \left( \eta_{\beta\nu} \sigma_{{ \rho}\mu} + \eta_{\beta\mu} \sigma_{{ \rho}\nu} \right) \Psi  +
 D_\beta  \bar\Psi  \left( \eta_{\alpha\nu}  \sigma_{{ \rho}\mu} + \eta_{\alpha\mu} \sigma_{{ \rho}\nu} \right) \Psi \right] \right. \nonumber\\
& - & \left.  \langle u^\alpha u^\beta\rangle  \left[
  \bar\Psi \left( \eta_{\beta\nu}  \sigma_{{ \rho}\mu} + \eta_{\beta\mu} \sigma_{{ \rho}\nu} \right) D_\alpha  \Psi  +
  \bar\Psi  \left( \eta_{\alpha\nu} \sigma_{{ \rho}\mu} + \eta_{\alpha\mu} \sigma_{{ \rho}\nu} \right)  D_\beta  \Psi
 \right]
  \right\}\nonumber\\ 
 & + & \frac{c_2}{4 m^2}  \left\{  \partial^\alpha \left[ \langle u_\alpha u_\mu \rangle  D_\nu ( \bar\Psi \Psi ) +\langle u_\alpha u_\nu \rangle  D_\mu ( \bar\Psi \Psi )  - \langle u_\mu u_\nu \rangle  D_\alpha ( \bar\Psi \Psi )\right] \right\}
  \nonumber\\
  &+& c_3 \, \bar\Psi \langle u_\mu u_\nu \rangle   \Psi 
 +\frac{i c_4}{8} \bar\Psi \left(  \sigma_{\nu\beta}\,\left[ u_\mu ,u^\beta\right] +\sigma_{\alpha\nu}\,\left[ u^\alpha ,u_\mu \right] +\sigma_{\mu\beta}\,\left[ u_\nu ,u^\beta\right] +\sigma_{\alpha\mu}\,\left[ u^\alpha ,u_\nu\right]  \right)  \Psi \nonumber\\
&+&  \frac{c_6}{8m} \bar\Psi \left( \sigma_{\nu\beta} F^+_{\mu\alpha} \eta^{\alpha\beta}+ \sigma_{\mu\beta} F^+_{\nu\alpha} \eta^{\alpha\beta} \right) \Psi 
+  \frac{c_7}{8m} \bar \Psi \left( \sigma_{\nu\beta} \langle F^+_{\mu\alpha}\rangle \eta^{\alpha\beta} + \sigma_{\mu\beta} \langle F^+_{\nu\alpha} \rangle\eta^{\alpha\beta}
  \right) \Psi 
  \nonumber\\
&+&  \frac{c_8}{4} (\eta_{\mu\nu}\partial^2 -\partial_\mu \partial_\nu ) \bar\Psi \Psi  
+ \frac{i c_9}{2 m} \, (\eta_{\mu\alpha} \eta_{\nu\beta} \partial^2 +\eta_{\mu\nu} \partial_\alpha \partial_\beta - \eta_{\mu\alpha} \partial_\nu \partial_\beta 
-\eta_{\nu\alpha} \partial_\mu \partial_\beta )  \nonumber\\
&\times&
\left( \bar\Psi \gamma^\alpha D^\beta  \Psi  - D^\beta  \bar\Psi \gamma^\alpha  \Psi +\bar\Psi \gamma^\beta D^\alpha  \Psi  - D^\alpha  \bar\Psi \gamma^\beta  \Psi \right)
  \nonumber\\
&-&  \eta_{\mu\nu}  \left[ 
\frac{1}{2} \, \bar\Psi \, i \gamma^\alpha D_\alpha \Psi -\frac{1}{2} \, D_\alpha \bar\Psi
\, i \gamma^\alpha\Psi  -m \bar\Psi\Psi +\frac{g_A}{2}\, \bar\Psi \gamma^\alpha \gamma_5 u_\alpha \Psi \right. \nonumber\\
&+& \left. c_1 \langle \chi_+\rangle  \bar\Psi  \Psi  - \frac{c_2}{8 m^2} \langle u_\alpha u_\beta\rangle  \left( \bar\Psi \left\{ D^\alpha, D^\beta\right\}  \Psi  
+
 \left\{ D^\alpha, D^\beta\right\}  \bar\Psi \Psi \right) + \frac{c_3}{2} \, \langle u_\alpha u^\alpha\rangle  \bar\Psi  \Psi
+ \frac{i c_4}{4} \, \bar\Psi\sigma^{\alpha\beta}\,\left[ u_\alpha ,u_\beta\right] \Psi \right. \nonumber\\
&+& \left. c_5 \bar\Psi\hat \chi_+ \Psi + \frac{c_6}{8m} \bar\Psi\sigma^{\alpha\beta} F^+_{\alpha\beta} \Psi + \frac{c_7}{8m} \bar \Psi \sigma^{\alpha\beta} \langle F^+_{\alpha\beta} \rangle \Psi 
 \right]  ,
\label{MEMT}
\end{eqnarray}
where
\begin{eqnarray}
D_\mu \Psi &=& \partial_\mu\Psi + \left(\Gamma_\mu  -i v_\mu^{(s)}\right)\Psi, \nonumber\\
D_\mu \bar\Psi &=& \partial_\mu\bar\Psi  - \bar\Psi \left( \Gamma_\mu  -i v_\mu^{(s)}\right) .
\label{CovDCh}
\end{eqnarray}

 The expression of Eq.~(\ref{MEMT}) can be used for calculations in the low-energy region of various matrix elements of the EMT (and of its various products with
scalar, pseudoscalar, vector and axial-vector quark currents)
between states containing one nucleon
and an arbitrary number of pions. Below we apply Eq.~(\ref{MEMT}) for calculations of the  one-loop corrections to GFFs of the nucleon up to fourth chiral order, and  
of the pion graviproduction tree-level amplitude in the leading and next-to-leading chiral orders.  The expression for the EMT of Eq.~(\ref{MEMT}) can be also applied to
variety of low-energy hadronic  processes induced by gravity (by EMT) and to calculations of various corrections to GFFs.

\begin{figure}[t]
\begin{center}
\epsfig{file=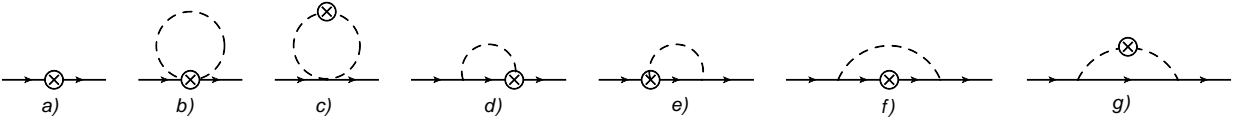,scale=0.99}
\caption{Tree and one-loop  diagrams contributing to the nucleon matrix element of the EMT. Dashed and solid lines correspond to pions and nucleons, respectively. The circles with crosses represent the EMT vertices. }
\label{EMT_nucleon}
\end{center}
\vspace{-5mm}
\end{figure}

\section{{One loop chiral corrections to nucleon gravitational form factors}}
\label{NFFs}

In this section we calculate the tree and one-loop contributions to the nucleon matrix element of the EMT.  The topologies of the corresponding diagrams  are shown in Fig.~\ref{EMT_nucleon}. 
Standard power counting rules apply to these diagrams \cite{Weinberg:1991um,Ecker:1994gg}, i.e. the pion lines count as of chiral order minus two, the nucleon lines have order minus one,  interaction vertices originating from the effective Lagrangian 
of order $N$ count also as of chiral order $N$ and the vertices generated by the { EMT} have the orders corresponding to the number of quark mass factors and derivatives acting on the pion fields, derivatives acting on the nucleon fields count as of chiral order 
zero. The momentum transfer between the initial and final nucleons also counts as of chiral order one, therefore in those terms of energy-momentum operator which contain full derivatives, these derivatives (although also acting on nucleon fields) count as of chiral order one.  
Integration over loop momenta is counted as of chiral order four. 

Since we are interested in the nucleon matrix element of order four in the chiral expansion, we need vertices with two nucleon lines, generated by  the EMT, up to order four. While we have obtained these vertices from the expression of Eq.~(\ref{MEMT}) 
for zeroth, first and second chiral orders, for the third and fourth order terms  we use a parameterisation as specified below.  Simple power counting arguments show that, because the pion-nucleon-nucleon 
vertices have at least chiral order one, for all one-loop diagrams except f)  
we only need vertices up to order two. Naively it seems that for diagrams of
topology f) we need also pion-nucleon-nucleon vertices of chiral order three,
because the nucleon-nucleon vertex originating from the  EMT starts with
chiral order zero. However more careful examination  reveals that the leading
order contribution of the diagram with the mentioned zeroth order  vertex is
exactly canceled by the nucleon wave function renormalization constant
multiplied by the tree order diagrams. Therefore the formally zeroth order vertex in effect 
starts contributing as a vertex of order one.  As a result of this the
diagram with the pion-nucleon-nucleon vertex of order three starts only
contributing at chiral order five. For this reason we do not consider such
diagrams in this work. Notice here that the above described power counting is 
realized in the results 
of our manifestly Lorentz-invariant calculations only after performing an appropriate renormalization.      

\medskip

The one-nucleon matrix element of the EMT is parameterised in terms of three form factors as follows \cite{Polyakov:2018zvc}:
\begin{eqnarray}
\langle p_f, s_f| T_{\mu\nu}| p_i,s_i \rangle &=& \bar u(p_f,s_f) \left[ A(t) \frac{P_\mu P_\nu}{m_N} + i J(t) \frac{P_\mu \sigma_{\nu\alpha} \Delta^\alpha + P_\nu \sigma_{\mu\alpha} \Delta^\alpha}{2 m_N}+ D(t) \frac{\Delta_\mu \Delta_\nu-\eta_{\mu\nu} q^2}{4 m_N} \right]  u(p_i,s_i) \,,
\label{EMTdef}
\end{eqnarray}
where $m_N$ is the physical mass of the nucleon, $(p_i,s_i)$ and $(p_f,s_f)$ are the momentum and polarization of the incoming and outgoing nucleons, respectively, and $P=(p_i+p_f)/2$, $\Delta=p_f-p_i$, { $t=\Delta^2$}.  

The tree-order diagrams up to chiral order four give the following contributions to the form factors:
\begin{eqnarray}
A_{\rm tree}(t) &=& 1 - \frac{2 c_9}{m_N} \, t + x_1 M_\pi^2 t +x_2 t^2 \,,\nonumber\\
J_{\rm tree}(t) &=& \frac{1}{2} - \frac{c_9}{m_N} \, t\,,\nonumber\\
D_{\rm tree}(t) &=&  c_8 m_N +  y_1 t+y_2 M_\pi^2 \,.
\label{temt}
\end{eqnarray}
where $c_8$ and $c_9$ terms are generated by the EMT of Eq.~(\ref{MEMT}), while $x_1$ and $x_2$ ($y_1$ and $y_2$) parameterize the tree-order contributions of the fourth (the third) chiral orders. 
 The  parameters   $x_i$ and $y_i$ are given as linear combinations of the coupling constants of the corresponding effective Lagrangians
 in the presence of the external gravitational field, derivation of which is beyond the scope of this work. 
 
In calculations of loop diagrams, shown in Fig.~\ref{EMT_nucleon}  we applied dimensional regularization (see, e.g., Ref.~\cite{Collins:1984xc}) and used the program
FeynCalc \cite{Mertig:1990an,Shtabovenko:2016sxi}. The One-loop expressions of the form factors are too large to be shown explicitly. Instead we give the corresponding expressions in chiral limit in the appendix.

We perform the renormalization of loop diagrams by applying the EOMS scheme \cite{Gegelia:1999gf,Fuchs:2003qc} with the renormalization scale $\mu=m_N$.  
Notice that the divergent pieces of one-loop contributions to $A(t)$, as well as to $J(t)$, with coefficients of chiral orders zero and two, vanish. 
On the other hand there is a power counting violating contribution to $A(t)$ 
given by $3 t g_A^2/(64 \pi ^2 F^2)$ which is absorbed into the renormalization of the coupling constant $c_9$ without affecting the power counting for $J(t)$ in which $c_9$ also gives a tree-order contribution.  
The coupling $c_8$ has to cancel the divergent part and the power counting violating piece of the one-loop contribution to $D(t)$ given by 
\begin{equation}
-\frac{g_A^2}{16 \pi ^2
   F^2} \left[ m_N^2  B_0\left(m_N^2,0,m_N^2\right)+ \text{A}_0\left(m_N^2\right)\right]-\frac{3 c_8 g_A^2 }{32 \pi ^2 F^2} \left(2 m_N^3-3 m_N
   \text{A}_0\left(m_N^2\right)\right) +\frac{3 c_9 g_A^2 m_N}{2 \pi ^2 F^2}  \text{A}_0\left(m_N^2\right)\,,
\label{divPPCV}
\end{equation}
where 
the loop integrals $A_0$ and $B_0$ are defined in the appendix.\footnote{Infinite renormalization of   $c_8$ implies that, while { being} a dimensionful coupling constant of an interaction of the gravitational 
and nucleon fields, it is only suppressed by hadronic { scale(s)}, because it receives corrections due to pion loops.}    

Adding the tree-order contribution of Eq.~(\ref{temt}) to the one-loop result we obtain the following expression for the $D$-term $D(0)$ expanded in powers of the pion mass:
\begin{eqnarray} 
\frac{D(0)}{m_N} & = & c_8 
+\frac{ g_A^2 }{16 \pi  F^2} \,M_{\pi } 
+\frac{ -3 g_A^2 /m_N+2 \left(-4
   c_1+c_2+2 c_3\right) }{8 \pi ^2 F^2} \, M_{\pi }^2 \ln\left( \frac{M_{\pi }}{m_N }\right) \nonumber\\
&+& \frac{\left(- g_A^2 \left(3 c_8 + 14/m_N\right)+8 c_3-16 c_1 \right)}{32 \pi ^2
   F^2}\ M_{\pi }^2 +\frac{y_2}{m_N}\ M_\pi^2 +{\cal O} (M_\pi^3)  \,.
\label{divPart}
\end{eqnarray}
We notice that our result is at variance with calculations of Diehl {\it et al.} \cite{Diehl:2006ya} done to the same chiral order using heavy baryon approach.
More specifically our coefficients of the non-analytical $\sim M_\pi^2 \ln (M_\pi/m_N)$ terms proportional to LECs $c_{2,3}$ are different from { those} of Ref.~\cite{Diehl:2006ya}.
For the remaining non-analytical contributions we found agreement with calculations
of  Ref.~\cite{Diehl:2006ya} and with other lower order calculations in Refs.~\cite{Belitsky:2002jp,Ando:2006sk,Dorati:2007bk,Moiseeva:2013qoa}.
Mentioned difference  might lead to the revision of
 the extrapolation of lattice data of nucleon GFFs to the physical point. 

\medskip

Next we define the slopes of GFFs by writing the form factors as:
\begin{eqnarray}
A(t) & = & 1+ s_A t +{\cal O}(t^2) \,,\nonumber\\
J(t) & = & \frac{1}{2}+ s_J t +{\cal O}(t^2) 
   \,,\nonumber\\
D(t) & = & D(0)+ s_D t +{\cal O}(t^2)  \,.
\label{defradii}
\end{eqnarray}
For the chiral expansion of the loop contributions to the slopes we obtain { (while the tree-order contributions are included in Eq.~(\ref{temt})\,) } 
\begin{eqnarray}
s_A & = &   
-\frac{7  g_A^2}{128 \pi  F^2 m_N}\ M_{\pi } + \frac{ \left(c_2 m_N-4 g_A^2\right)}{16 \pi ^2 F^2 m_N^2}\ M_{\pi }^2 \ln \left(\frac{M_{\pi }}{m_N}\right)-\frac{3
    g_A^2 \left(2 c_9 m_N+1\right)}{32 \pi ^2 F^2 m_N^2}\ M_{\pi }^2  +{\cal O} (M_\pi^3)  \,,\nonumber\\
   s_J & = & 
  -\frac{g_A^2 }{32 \pi ^2
   F^2}\ \ln\left( \frac{M_{\pi }}{m_N}\right)+ \frac{g_A^2 \left(4 c_9 m_N-5\right)}{64 \pi ^2 F^2} 
  +\frac{7  g_A^2}{128 \pi  F^2 m_N}\ M_{\pi }  +{\cal O} (M_\pi^2)  
   \,,\nonumber\\
s_D & = & -\frac{g_A^2 m_N}{40 \pi  F^2 }\ \frac{1}{M_\pi} -\frac{ \left(5 g_A^2+4 \left(c_2+5 c_3\right) m_N\right)}{80 \pi ^2
   F^2} \ \ln \left(\frac{M_{\pi }}{m_N}\right)+ { \frac{g_A^2 \left(24+ (15 c_8+40 c_9) m_N \right)}{480 \pi ^2 F^2} } \nonumber\\
   &+& \frac{\left(4 c_1-c_2-7 c_3\right) m_N}{40 \pi ^2 F^2} +{\cal O} (M_\pi) \,.
\label{radii}
\end{eqnarray}
Again the non-analytic terms  $\sim c_{2,3} M_\pi^2\ln (M_\pi/m_N)$ in our calculation differ from those of Ref.~\cite{Diehl:2006ya}.
For other non-analytical contributions we find agreement with calculations
of  Ref.~\cite{Diehl:2006ya} and with other lower order calculations in Refs.~\cite{Belitsky:2002jp,Ando:2006sk,Dorati:2007bk,Moiseeva:2013qoa}.

\subsection{Large distance  asymptotics of the energy, spin, pressure and shear force distributions}
\label{sec:largeR}

The GFFs of the nucleon  $A(t), J(t)$ and $D(t)$ can be related to the energy and spin densities as \cite{Polyakov:2002yz,Polyakov:2018zvc}:
\begin{eqnarray}\label{EQ:staticEMT-T00}
	\rho_{E}(r)&=&m_N \int \frac{d^3 {\bf\Delta}}{(2\pi)^3}\ 
	e^{{-i}{\bf  r\Delta}} \,\biggl[
        A(-{\bf  \Delta}^2) + \frac{{\bf  \Delta}^2}{4m_N^2}\bigl[A(-{\bf  \Delta}^2)-2J(-{\bf  \Delta}^2)+D(-{\bf  \Delta}^2)\bigr]\biggr]\,,\\
 \rho_{J}(r)&=&  \int \frac{d^3 {\bf \Delta}}{(2\pi)^3}\ 
	e^{{-i}{\bf  r\Delta}} \,
       \left[ J(-{\bf  \Delta}^2)+\frac 23 {\bf  \Delta}^2 \frac{dJ(-{\bf  \Delta}^2)}{d {\bf  \Delta}^2}\right] \,.   
\end{eqnarray}
The vector field of the spin distribution 
in the polarised nucleon has the form $J^i({\bf r}, {\bf s}) =\frac 32 \left(\delta^{ik} -\frac{r^ir^k}{r^2}\right) s^k \rho_J(r)$ \cite{Lorce:2017wkb,Schweitzer:2019kkd}.
The distribution of the pressure $p(r)$ and shear force $s(r)$ are obtained through \cite{Polyakov:2002yz,Polyakov:2018zvc}:

\begin{eqnarray}
\label{Eq:relationSPD}
	s(r)= -\frac{1}{4 m_N}\ r \frac{d}{dr} \frac{1}{r} \frac{d}{dr}
	{\widetilde{D}(r)}, \quad
	p(r)=\frac{1}{6 m_N} \frac{1}{r^2}\frac{d}{dr} r^2\frac{d}{dr}
 	{\widetilde{D}(r)}, \quad
	{\widetilde{D}(r)=}
	\int {\frac{d^3{\bf  \Delta}}{(2\pi)^3}}\ e^{{-i} {\bf  \Delta r}}\ D(-{\bf  \Delta}^2).
\end{eqnarray}
The large distance power-like behavior  of the distributions in  the parametrically wide region $1/\Lambda_{\rm strong}\ll r\ll 1/M_\pi$  
is governed by the singularities of GFFs at  $t=0$, i.e. by the non-analytical  terms of the GFFs in the chiral limit. From our loop calculations we can easily obtain 
the following small $t$ behavior of the GFFs in the chiral limit to the accuracy of our calculations:
\begin{eqnarray}
	A (t) &=& { 1-\frac{2 c_9 }{m_N}\ t  +\frac{3 g_A^2  }{512 F^2 m_N} (-t)^{\frac 32} -\frac{ \left(c_2 m_N-10 g_A^2\right)}{320 \pi ^2 F^2 m_N^2} \ t^2 \ln\left( \frac{-t}{m_N^2}\right) -\frac{ \left(25 g_A^2 \left(12 c_9 m_N-7\right)-62 c_2 m_N\right)}{9600 \pi ^2 F^2 m_N^2} t^2 
} + {  O(t^{\frac{5}{2}}) }  \,, \nonumber\\
J (t) &=& {  \frac{1}{2}-\frac{c_9 }{m_N}\ t -\frac{ g_A^2 }{64 \pi ^2 F^2}\ t \ln \left(\frac{-t}{m_N^2}\right) + { \frac{g_A^2 \left(12 c_9 m_N-7\right)}{192 \pi ^2 F^2}\ t  }  -\frac{3 g_A^2  }{512 F^2 m_N} }
   (-t)^{\frac 32} + { O(t^2)  } 
    \,,\nonumber\\
    D (t)    &= & { m_N c_8 +\frac{3
    g_A^2 m_N}{128 F^2}\sqrt{-t} -\frac{\left(5 g_A^2+4 \left(c_2+5 c_3\right) m_N\right)}{160 \pi ^2 F^2}\  t
   \ln \left(\frac{ - t}{m_N^2} \right) } \nonumber\\
   &+& {\frac{\left(5 g_A^2 \left(40 c_9 m_N+15 c_8 m_N+28\right)+94 c_2 m_N+200 c_3 m_N\right)}{2400 \pi ^2 F^2} \ t
   + { O(t^{\frac{3}{2}})} \,.}
\label{FFsChL1}
\end{eqnarray}
Performing 3D Fourier transformation of these expressions we obtain the  large distance behavior of the spatial  distributions in  the parametrically wide region $1/\Lambda_{\rm strong}\ll r\ll 1/M_\pi$:
\begin{eqnarray}
\label{eq:ElargeR}
\rho_E(r)&=& \frac{9
    g_A^2 }{64 \pi^2 F^2} \frac{1}{r^6} { -\frac{ 3 \left(10 g_A^2/m_N+ \left(c_2+10 c_3\right) \right)}{16 \pi ^3 F^2}\ \frac{1}{r^7}}+O\left(\frac{1}{r^8}\right) \,, \\
\rho_J(r)&=& {  \frac{ 5 g_A^2  }{64 \pi ^3 F^2} \frac{1}{r^5}  -\frac{9 g_A^2  }{64 \pi^2 F^2 m_N} \frac{1}{r^6} +O\left(\frac{1}{r^7}\right)} \,,\\
\label{eq:DlargeR}
\widetilde{D}(r)&=&-\frac{3
    g_A^2 m_N}{128 \pi^2 F^2} \frac{1}{r^4}  + \frac{ 3 \left(5 g_A^2+4 \left(c_2+5 c_3\right) m_N\right)}{160 \pi ^3 F^2}\ \frac{1}{r^5}+O\left(\frac{1}{r^6}\right) \,.
    \label{Dtilde}
\end{eqnarray}
{ Using Eq.~(\ref{Dtilde}) in Eq.~(\ref{Eq:relationSPD}) we} obtain the large distance behavior 
 of the pressure and shear force distributions:
\begin{eqnarray}
\nonumber
p(r)&=&-\frac{3
    g_A^2 }{64 \pi^2 F^2} \frac{1}{r^6}{ +\frac{  \left(5 g_A^2/m_N+4 \left(c_2+5 c_3\right) \right)}{16 \pi ^3 F^2}\ \frac{1}{r^7}} +O\left(\frac{1}{r^8}\right), \\
\label{eq:SPlargeR}
 s(r)&=&  \frac{9
    g_A^2 }{64 \pi^2 F^2} \frac{1}{r^6}  { - \frac{ 21 \left(5 g_A^2/m_N+4 \left(c_2+5 c_3\right) \right)}{128 \pi ^3 F^2}\ \frac{1}{r^7}} +O\left(\frac{1}{r^8}\right)\,.
 \label{MaxX}    
\end{eqnarray}
The leading  terms ($\sim 1/r^6$) { in Eq.~(\ref{MaxX}) } have been obtained for the first time in Ref.~\cite{Goeke:2007fp} in the framework of the soliton picture of the nucleon.
The obtained large distance asymptotics can be useful for the derivation of various inequalities for the bound states of various quarkonia with the nucleon
in the hadro-quarkonium picture of the exotic pentaquarks with hidden heavy quarks content, see, e.g., Ref.~\cite{Perevalova:2016dln}. Also it can be useful 
for the analysis of lattice data on GFFs of the nucleon and for deriving general constraints on the GFFs. To illustrate the latter point we note
that the large distance behavior of the energy density,   given by Eq.~(\ref{eq:ElargeR}), and of pressure and the shear force distributions,  specified in Eq.~(\ref{eq:SPlargeR}),
satisfy the general stability conditions - $\rho_E(r)>0$ and $\frac 23 s(r)+p(r)>0$, see discussion in Ref.~\cite{Polyakov:2018zvc}.

 Furthermore with help of expression for $J(t)$ in Eq.~(\ref{FFsChL1})  we can obtain  large impact-parameter 
behavior of the distributions of Belinfante-improved 
total angular momentum. The latter is defined as \cite{Lorce:2017wkb}:
\begin{eqnarray}
\langle J_{\rm Bel} \rangle (b_\perp)&=& -\frac{1}{2} b_\perp \frac{\partial}{\partial b_\perp}{\int \frac{d^2 {\bf \Delta_\perp}}{(2\pi)^2}\ 
	e^{{-i}{\bf  b_\perp \Delta_\perp}} \,
       J(-{\bf  \Delta_\perp}^2)\,.   }
\end{eqnarray}
Performing the 2D Fourier transformation we obtain the large $b_\perp$ asymptotics of $\langle J_{\rm Bel} \rangle (b_\perp)$ as:
\begin{eqnarray}
\langle J_{\rm Bel} \rangle (b_\perp)= \frac{g_A^2}{16\pi^3 F^2} \frac{1}{b_\perp^4}- \frac{135 g_A^2}{2048 \pi F^2 m_N} \frac{1}{b_\perp^5}+
O\left(\frac{1}{b_\perp^6}\right)\,.   
\end{eqnarray}
This model-independent asymptotics is valid in the parametrically wide region $1/\Lambda_{\rm strong}\ll b_\perp \ll 1/M_\pi$ and can be used for
derivation of  model-independent constraints for total angular momentum distribution in the nucleon.

\section{Pion graviproduction}
\label{ppr}

The effective action  of Eq.~(\ref{MAction})  obtained  from the effective Lagrangian is universal and, hence, can be applied to a wide range of low-energy processes induced by gravity. 
As an example of an application of the action  of Eq.~(\ref{MAction})  we consider in this section
 the tree-order amplitude of the one-pion production in a gravitational field close to threshold, where the chiral expansion of this quantity makes sense.  
 The pion graviproduction is relevant not only for  hadronic reactions in  strong gravitational fields, but it can also be measured in hard exclusive processes \cite{Polyakov:1998sz,Guichon:2003ah,Polyakov:2006dd,Kivel:2004bb}.
 
 \begin{figure}[t]
\begin{center}
\epsfig{file=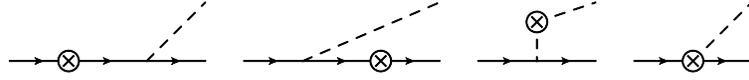,scale=0.6}
\caption{Tree-order  diagrams contributing to the pion  graviproduction.   Dashed and solid lines correspond to pions and nucleons, respectively. The circles with crosses represent the EMT vertices. }
\label{Pion_P}
\end{center}
\vspace{-5mm}
\end{figure}

The full tree-order amplitude of  the one pion production ${\cal M}^{a, \mu\nu}=\langle \pi^a(k) N(p_f) | T^{\mu\nu}(0)  | N(p_i)\rangle$ of leading and next-to-leading chiral orders  is given   in Eq.~(\ref{Ttree}) of the appendix - the corresponding diagrams are shown in Fig.~\ref{Pion_P}.  
Here we give the expression for the amplitude at threshold which  can be parameterised in terms of GFFs for the $1/2^+\to 1/2^-$ transition. The corresponding form
factors were introduced first in Ref.~\cite{Kobzarev:1962wt}  for the case of equal masses of the final and initial states, for different masses they were  considered in Ref.~\cite{MVPAT}.
Following the latter reference we obtain the threshold amplitude of the pion graviproduction in the following form: 
\begin{eqnarray}
-i\ {\cal M}_{\rm tree}^{a, \mu\nu} &=& F_1({  \Delta^2}) \, \bar u(p_f) \frac{\tau^a}{2}\left\{ \,  \Delta^2 P^\mu P^\nu - (m_*^2-m_N^2) \Delta^{\{\mu} P^{\nu\}} + \frac{(m_*^2-m_N^2)^2}{4}\,g^{\mu\nu} 
 \right\} \gamma^5 u(p_i)\nonumber\\
 &+& F_2({  \Delta^2}) \, \bar u(p_f ) \frac{\tau^a}{2}\left\{ \,  \Delta^2 \gamma^{\{\mu} P^{\nu\}}  - (m_*+m_N) \Delta^{\{\mu} P^{\nu\}} - \frac{(m_*^2-m_N^2)}{2}\, \left( \gamma^{\{\mu} \Delta^{\nu\}}  -(m_*+m_N)g^{\mu\nu} \right) 
 \right\} \gamma^5 u(p_i ) \nonumber\\
 &+& F_3({  \Delta^2}) \, \left\{ \,  \Delta^\mu \Delta^\nu - g^{\mu\nu} \Delta^2
 \right\} \bar u(p_f) \frac{\tau^a}{2} \gamma^5 u(p_i)\,, 
 \end{eqnarray}
 where $\Delta =p_f+k-p_i $, $P=p_i+\frac{\Delta}{2}$, $m_*=m_N+M_\pi$, and the symmetrization is defined as $X^{\{\mu} Y^{\nu\}}=(X^\mu Y^\nu+ X^\nu Y^\mu)/2$. 
 
Our tree order calculation gives:
 \begin{eqnarray}
	 F_1({  \Delta^2}) & =& \frac{8 \,c_9 g_A m_N}{F { m_*} \left(m_N
   \left(m_*+m_N\right)- {  \Delta^2} \right)}+\frac{4  g_A m_N \left(m_N-M_{\pi
   }\right)}{F \left(m_N \left(m_*+m_N\right)- {  \Delta^2} \right) \left(M_{\pi
   }^2 \left(m_*+m_N\right)- { \Delta^2} m_N\right)}\,,\nonumber\\
 F_2({  \Delta^2}) &=&  
 \frac{2   g_A m_N}{F \left({  m_*+m_N} \right) \left(m_N
 \left({ {m_*}}+m_N\right)-{  \Delta^2}\right)}-\frac{2 \,c_9  g_A \left(M_{\pi }
 m_N \left({ m_*}+m_N\right){}^2+{  \Delta^2} \left(2 m_N^2-M_{\pi }^2\right)\right)}{F
	 { m_*} m_N \left({ m_*+ m_N} \right) \left(m_N
 \left({ m_*}+m_N\right)-{  \Delta^2}\right)}
  \,,\nonumber\\
  F_3({  \Delta^2}) &=& -\frac{ c_9 M_{\pi } g_A \left(-M_{\pi }^2 m_N+4 m_N^3+M_{\pi } { \Delta^2}\right)}{F
   m_* m_N \left(m_N \left(m_*+m_N\right)-{  \Delta^2}\right)}-\frac{c_8 m_* g_A
   \left(2 m_N \left(m_*+m_N\right)-{  \Delta^2}\right)}{2 F \left(m_*+m_N\right)
   \left(m_N \left(m_*+m_N\right)-{  \Delta^2}\right)} \nonumber\\
   &+&  \frac{ M_{\pi } g_A m_N \left(-4
   m_N^2+M_{\pi }^2+{  \Delta^2}\right)}{F \left(m_N \left(m_*+m_N\right)-{  \Delta^2}\right)
   \left(M_{\pi }^2 \left(m_*+m_N\right)-{  \Delta^2} m_N\right)}\,.
\label{Ttree}
\end{eqnarray}
In the momentum transfer range ${  \Delta^2}\sim M_\pi^2\sim \varepsilon^2\to 0$ the above form factors scale as $F_1({  \Delta^2})\sim 1/\varepsilon^2$,
$F_2({  \Delta^2})\sim \varepsilon^0$, and $F_3({  \Delta^2})\sim 1/\varepsilon$.

 Our amplitude of the pion graviproduction  (see the above equation and Eq.~(\ref{Ttree}) in the appendix) depends on the new LECs $c_8$ and $c_9$. Therefore
the measurements of the pion graviproduction process can be used as an additional source of information on these LECs,  and also for studying   the
 applicability  of the chiral effective field theory to reactions induced by  gravitational interactions. The pion  graviproduction can be accessed in hard exclusive processes like the 
non-diagonal DVCS ($\gamma^*+ N\to \gamma+(\pi N)$)  \cite{Polyakov:1998sz,Guichon:2003ah,Polyakov:2006dd,Kivel:2004bb}. The corresponding
measurements are planned by the CLAS12 collaboration at JLab (USA) \cite{privatcomm}.

\section{Summary}
\label{conclusions}

In the current work we presented the effective chiral Lagrangian of pions and nucleons   up to the second chiral order in the presence of external gravitational field. 
We derived the corresponding energy-momentum tensor of pions and nucleons (with external scalar, pseudoscalar, vector and  axial-vector quark currents included)
in flat space-time.  Next we calculated the one-loop contributions to 
the one-nucleon matrix element of the energy-momentum tensor at fourth chiral order  and extracted the corresponding 
contributions to gravitational form factors of the nucleon. 
 To renormalize the loop diagrams we applied the EOMS renormalization scheme of Refs.~\cite{Gegelia:1999gf,Fuchs:2003qc}.
For the tree contributions of the first and the second order we obtained the Feynman rules from the corresponding expressions of the energy-momentum tensor while for the third and fourth orders we used parametrizations in most general form. 
While the coefficients of these parametrizations can be expressed as linear combinations of the coupling constants of the effective Lagrangians of the corresponding chiral orders in the presence 
of an external gravitational field, derivation of these Lagrangians and the corresponding energy-momentum tensors is beyond the scope of this work. As  the obtained expressions for the gravitational form factors of the nucleon, 
defined by the matrix element of the energy-momentum tensor, are too large to be given explicitly, in the appendix 
we quote them in the chiral limit. We calculated the chiral expansion of the $D(0)$ and slope parameters for all GFFs to the fourth order of the chiral expansion.
 Our results for non-analytical contributions of the type $\sim c_{2,3} M_\pi^2 \ln (M_\pi/m_N)$ differ from those of the previous fourth order calculations of Ref.~\cite{Diehl:2006ya}.
For the remaining non-analytical contributions we found agreement with the calculations
of  Ref.~\cite{Diehl:2006ya} and with other lower order calculations in Refs.~\cite{Belitsky:2002jp,Ando:2006sk,Dorati:2007bk,Moiseeva:2013qoa}.  The 
difference of our results with those in Ref.~\cite{Diehl:2006ya} was discussed with the authors of Ref.~\cite{Diehl:2006ya} and they agreed with our calculations.
 It is very important to check how strong the new results  affect the chiral extrapolation of the lattice data on nucleon GFFs obtained
in the past. However this is beyond the scope of the current work.

Furthermore we calculated the leading- and next-to-leading order tree contributions to the amplitude of the pion graviproduction. 
The process of the pion graviproduction provides an  additional independent (to gravitational form factors) source of information on the new LECs $c_{8,9}$.
 It also offers a test of applicability of chiral perturbation theory to gravity-induced low-energy processes. Possibility of measurements of the pion graviproduction in hard exclusive processes 
 has been discussed in Refs.~\cite{Polyakov:1998sz,Guichon:2003ah,Polyakov:2006dd,Kivel:2004bb}.  
Such kind of experiments are planned by the CLAS12 collaboration at JLab (USA) \cite{privatcomm}. The effective action of Eq.~(\ref{MAction}) obtained here
provides a systematic tool of analyzing the data on these processes.

\section*{Acknowledgments}

  We are sincerely grateful to Alexander Manashov for sharing details of his calculations and appreciate very much his help in clarifying the difference of our results with that
of  Ref.~\cite{Diehl:2006ya}.
We thank U.-G. Mei\ss ner for the comments on the manuscript, and  MVP acknowledges helpful discussions with V.~Burkert, S.~Diehl, and K.~Joo about feasibility  of measuring the non-diagonal DVCS processes.
This work was supported in part by BMBF (Grant No. 05P18PCFP1),  Georgian Shota Rustaveli National
Science Foundation (Grant No. FR17-354), and  by the Sino - German CRC 110 ``Symmetries and the Emergence of Structure in QCD".

\appendix

\section{Definition of loop integrals}

The one-loop integrals appearing in expressions of our quoted results are defined as follows:
\begin{eqnarray}
A_0(m^2)&=&\frac{(2\pi)^{4-n}}{i\pi^2} \int\frac{d^n k}{k^2-m^2+i \epsilon}\,,\nonumber\\
B_0(p^2,m_1^2,m_2^2)&=&\frac{(2\pi)^{4-n}}{i\pi^2} \int\frac{d^nk}{[k^2-m_1^2+i \epsilon] [(p+k)^2-m_2^2+i \epsilon]}\,, \nonumber\\
C_0(p_1^2,p_2^2,p^2_{12},m_1^2,m_2^2,m_3^2)&=&\frac{(2\pi)^{4-n}}{i\pi^2} \int\frac{d^nk}{[k^2-m_1^2+i \epsilon] [(p_1+k)^2-m_2^2+i \epsilon] [(p_1+p_2+k)^2-m_ 3^2+i \epsilon]}\,,
\label{ints}
\end{eqnarray}
 with $p_{12}=p_1+p_2$.
  For the reduction of tensor loop integrals to scalar ones we apply the formulae specified in Ref.~\cite{Denner:2005nn}  
while for the expansion in terms of kinematical invariants we use Ref.~\cite{Devaraj:1997es}. 

\section{One-loop expressions for form factors in chiral limit}

Renormalized expressions of form factors in chiral limit for $\mu=m_N$:
\begin{eqnarray}
A (t) &=& 1 -\frac{2 c_9 t}{m_N} +
\frac{t g_A^2 }{64 \pi ^2 F^2
   \left(4 m_N^2-t\right){}^3} \biggl[ -48 t m_N^4 \left(m_N^2+t\right)
   \text{C}_0\left(m_N^2,m_N^2,t,0,m_N^2,0\right) 
   \nonumber\\
   && -2 m_N^2 \left(t-4
   m_N^2\right){}^2 B_0\left(t,m_N^2,m_N^2\right)-4 t m_N^2 B_0(t,0,0) \left(26
   m_N^2+t\right)-42 t^2 m_N^2+248 t m_N^4-32 m_N^6+3 t^3\biggr]
   \nonumber\\
   &&
   +\frac{c_9 t^2 g_A^2 m_N
   \left(B_0\left(t,m_N^2,m_N^2\right)-1\right)}{16 \pi ^2 F^2 \left(4
   m_N^2-t\right)}+\frac{c_2 t^2 \left(15 B_0(t,0,0)+16\right)}{4800 \pi ^2 F^2
   m_N}
\,, \nonumber\\
   J (t) &=& \frac{1}{2} -\frac{c_9 t}{m_N} +\frac{g_A^2  }{128 \pi ^2 F^2 \left(t-4 m_N^2\right){}^2}
   \biggl[ 24 t^2 m_N^4 \text{C}_0\left(m_N^2,m_N^2,t,0,m_N^2,0\right)+4 t
   m_N^2 B_0(t,0,0) \left(8 m_N^2+t\right)
  \nonumber\\
   &&  
   +2 \left(t m_N-4 m_N^3\right){}^2
   B_0\left(t,m_N^2,m_N^2\right)+30 t^2 m_N^2-128 t m_N^4+32 m_N^6-3
   t^3\biggr]
   -\frac{c_9 t g_A^2 m_N
   \left(B_0\left(t,m_N^2,m_N^2\right)-1\right)}{32 \pi ^2 F^2}
    \,,\nonumber\\
D (t) & = & m_N c_8 +\frac{g_A^2 m_N^2}{48 \pi ^2 F^2 t \left(t-4
   m_N^2\right){}^2}  \biggl[ 6 m_N^2 \left(3 t^2 \left(2 m_N^2-t\right)
   \text{C}_0\left(m_N^2,m_N^2,t,0,m_N^2,0\right)-2 \left(t-4 m_N^2\right){}^2
   B_0\left(t,m_N^2,m_N^2\right)\right)
   \nonumber\\
   &&+6 t^2 B_0(t,0,0) \left(m_N^2-t\right)-22
   t^2 m_N^2+128 t m_N^4-192 m_N^6+5 t^3\biggr]
   +\frac{3 c_8 g_A^2 m_N^3
   \left(B_0\left(t,m_N^2,m_N^2\right)+1\right)}{16 \pi ^2 F^2}
    \nonumber\\
   &&
   +\frac{c_9 g_A^2
   m_N^3 \left(B_0\left(t,m_N^2,m_N^2\right)+1\right)}{2 \pi ^2 F^2} +\frac{c_2 t \left(15
   B_0(t,0,0) \left(8 m_N^2+3 t\right)+68 m_N^2-42 t\right)}{4800 \pi ^2 F^2 m_N} 
    \nonumber\\
   &&
    +\frac{c_3 t
   m_N \left(3 B_0(t,0,0)-1\right)}{24 \pi ^2 F^2}
   \,.
   \label{FFsChL}
\end{eqnarray}

\section{Tree-order amplitude of the one-pion production}

Tree-order amplitude of the pion production  ${\cal M}^{a, \mu\nu}=\langle \pi^a(k) N(p_f) | T^{\mu\nu}(0)  | N(p_i)\rangle$ has the following form:
\begin{eqnarray}
-i\ {\cal M}_{\rm tree}^{a, \mu\nu} &=&  \bar u(p_f, s_f ) \frac{\tau^a}{2}\Biggl\{ \, \frac{g_A }{2 F \left(m_N^2-s\right)}  \left(c_8 m_N \left(q^2 g^{\mu \nu }-q^{\mu } q^{\nu }\right)
   \kslash \gamma ^5+\left(m_N-2 c_9 q^2\right)
   \left(\left(2\ { p_i ^{\nu }}+q^{\nu }\right) \kslash \gamma ^{\mu } \gamma ^5
   \right. \right.  \nonumber\\ 
   &+ & \left. \left.  \left(2\ { p_i ^{\mu }}+q^{\mu
   }\right) \kslash \gamma ^{\nu } \gamma
   ^5\right)\right)  +\frac{g_A }{2
   F \left(m_N^2-u\right)}  \left(c_8 m_N \left(q^2 g^{\mu \nu }-q^{\mu } q^{\nu }\right)
   \kslash \gamma ^5+\left(2 c_9 q^2-m_N\right)
   \left(\left(2  p_f^{\nu }-q^{\nu }\right) \gamma ^{\mu }\kslash  \gamma ^5 
  \right. \right.  \nonumber\\ 
   &+ & \left. \left.  
   \left(2 p_f^{\mu }-q^{\mu }\right)
   \gamma ^{\nu } \kslash \gamma ^5\right)\right)  +\frac{  g_A m_N }{F \left(M_\pi^2-t\right)} \,  \gamma ^5 \left(q^2 g^{\mu \nu }+
   4 \text{\textit{k}}^{\mu } \text{\textit{k}}^{\nu }-2 q^{\nu } \text{\textit{k}}^{\mu }-2 q^{\mu }
   \text{\textit{k}}^{\nu }\right) \nonumber\\
   &-& \frac{c_9 g_A }{2 F
   m_N} \left(2
   \left(q^2-t\right) g^{\mu \nu } \kslash\gamma
   ^5-2 \left(q^{\nu } \text{\textit{k}}^{\mu }+q^{\mu }
   \text{\textit{k}}^{\nu }\right) \left(\kslash\gamma ^5-2 \gamma ^5 m_N\right)+\gamma ^{\mu }\gamma
   ^5 \left(2 q^2 \text{\textit{k}}^{\nu }+q^{\nu }
   \left(t-q^2\right)\right)
   \right. \nonumber\\
   &+& \left. 
   \gamma ^{\nu }\gamma ^5 \left(2 q^2
   \text{\textit{k}}^{\mu }+q^{\mu } \left(t-q^2\right)\right)\right) 
   +\frac{c_9 M_{\pi }^2 g_A }{2 F m_N}  \left(-2 g^{\mu \nu } \kslash \gamma ^5+\gamma ^{\mu }\gamma ^5 q^{\nu }+\gamma ^{\nu
   }\gamma ^5 q^{\mu }\right)  
   \nonumber\\
 &+& \frac{ c_8 g_A }{2 F}  \gamma ^5 \left(q^2 g^{\mu \nu }-q^{\mu } q^{\nu
   }\right)+\frac{ g_A m_N }{F} \, \gamma ^5 g^{\mu \nu } 
    \Biggr\} u(p_i, s_i)\,,
\label{Ttree}
\end{eqnarray}
where $q=p_f+k-p_i$  and the Mandelstam variables $s,t,u$ are defined as: $s=(p_f+k)^2, t=(p_f-p_i)^2, u=(k-p_i)^2 $.
 One can easily check that the obtained amplitude is explicitly transverse, i.e. $q_\mu{\cal M}^{a, \mu\nu}=q_\nu{\cal M}^{a, \mu\nu}=0$ as it follows from conservation of EMT.

\end{document}